\documentclass[showpacs,aps,prl,twocolumn,superscriptaddress]{revtex4-1}
\usepackage{graphicx} 
\usepackage{dcolumn}
\usepackage{bm}
\usepackage{array}
\usepackage{amssymb,amsmath}
\usepackage{epstopdf}
\usepackage{color}
\usepackage{mathrsfs}
\def\etal{$\it{et~al.}$}
\newcolumntype{C}[1]{>{\centering\arraybackslash}p{#1}}

\begin{document}
\title{Unraveling the Topological Phase of ZrTe$_5$ via Magneto-infrared Spectroscopy}
\author{Y. Jiang}
\affiliation{National High Magnetic Field Laboratory, Tallahassee, Florida 32310, USA}
\author{J. Wang}
\affiliation{School of Physics, Georgia Institute of Technology, Atlanta, Georgia 30332, USA}
\affiliation{State Key Laboratory for Artificial Microstructure and Mesoscopic Physics, Peking University, Beijing 100871, China}
\author{T. Zhao}
\affiliation{School of Physics, Georgia Institute of Technology, Atlanta, Georgia 30332, USA}
\author{Z. L. Dun}
\affiliation{School of Physics, Georgia Institute of Technology, Atlanta, Georgia 30332, USA}
\author{Q. Huang}
\affiliation{Department of Physics and Astronomy, University of Tennessee, Knoxville, Tennessee 37996, USA}
\author{X. S. Wu}
\affiliation{State Key Laboratory for Artificial Microstructure and Mesoscopic Physics, Peking University, Beijing 100871, China}
\author{M. Mourigal}
\affiliation{School of Physics, Georgia Institute of Technology, Atlanta, Georgia 30332, USA}
\author{H. D. Zhou}
\affiliation{Department of Physics and Astronomy, University of Tennessee, Knoxville, Tennessee 37996, USA}
\author{W. Pan}
\affiliation{Quantum and Electronic Materials Department, Sandia National Laboratories, Livermore, California 94551, USA}
\author{M. Ozerov}
\affiliation{National High Magnetic Field Laboratory, Tallahassee, Florida 32310, USA}
\author{D. Smirnov}
\affiliation{National High Magnetic Field Laboratory, Tallahassee, Florida 32310, USA}
\author{Z. Jiang}
\email{zhigang.jiang@physics.gatech.edu}
\affiliation{School of Physics, Georgia Institute of Technology, Atlanta, Georgia 30332, USA}
\date{\today}

\begin{abstract}
For materials near the phase boundary between weak and strong topological insulators (TIs), their band topology depends on the band alignment, with the inverted (normal) band corresponding to the strong (weak) TI phase. Here, taking the anisotropic transition-metal pentatelluride ZrTe$_5$ as an example, we show that the band inversion manifests itself as a second extremum (band gap) in the layer stacking direction, which can be probed experimentally via magneto-infrared spectroscopy. Specifically, we find that the band anisotropy of ZrTe$_5$ features a slow dispersion in the layer stacking direction, along with an additional set of optical transitions from a band gap next to the Brillouin zone center. Our work identifies ZrTe$_5$ as a strong TI at liquid helium temperature and provides a new perspective in determining band inversion in layered topological materials.
\end{abstract}


\maketitle

Narrow-gap semiconductors and semimetals have regained broad interests in the past decade, as they host a rich variety of topological materials including topological insulators (TIs) and semimetals \cite{TI_review_0,TI_review,Vafek,Burkov,Armitage}. The low-energy electronic structure of such materials usually exhibits mixing characters of both linear band (LB, $E \propto k$, where $k$ is the wave vector) and parabolic band (PB, $E \propto k^2$), if higher-order terms are neglected. This concept is well reflected in the effective TI model \cite{BHZ_Zhang1,BHZ_Zhang2}, which has been proven successful in describing many topological material systems such as HgTe quantum wells \cite{BHZ_Molenkamp,BHZ_Buttner,BHZ_Smirnov}, Bi$_2$Se$_3$ \cite{BHZ_Zhang1,BHZ_Orlita}, alkali pnictides $A_3$Bi ($A=$Na, K, Rb) \cite{BHZ_Fang_1}, Cd$_3$As$_2$ \cite{BHZ_Fang_2}, Pb$_{1-x}$Sn$_x$Se \cite{PbSnSe_1,PbSnSe_2}, and transition-metal pentatelluride ZrTe$_5$ \cite{Refl_NLW,YJ_ZT5}.

The rising interest in ZrTe$_5$ is due to the theoretical prediction of a room-temperature quantum spin Hall insulator phase in its monolayer limit \cite{Thoery1_FZ} and the experimental observation of the chiral magnetic effect \cite{Arpes0_GDG}, anomalous Hall effect \cite{Ong}, and three-dimensional (3D) quantum Hall effect \cite{3DQH} in bulk material. However, because of the delicate dependence of its band topology on the lattice constants \cite{Thoery1_FZ,NewTheory_Zhou}, there has not yet been a consensus on the bulk topological phase of ZrTe$_5$ from experiments \cite{Arpes0_GDG,Refl1_NLW,ARST_SHP,STM_XQK,Arpes4_AC,Chen_PNAS}, especially with several recent contradicting temperature-dependent studies \cite{Arpes1_MG,Arpes2_XJZ,Arpes5_YLC,Temp1_ZT5,Strain_ZT5,Temp2_ZT5}. In these studies, the non-trivial topological phase is either probed through its surface states or relying on the transition behavior of certain indirect parameters such as conductivity. Alternatively, one could also seek to probe the PB component of ZrTe$_5$, which is associated with band inversion and thus provide direct evidence of the band topology without changing external parameters such as temperature and strain. However, this direct approach has not been reported to date.

In this Letter, we show that the LB and PB components of the electronic structure of ZrTe$_5$ can be determined using magneto-infrared (magneto-IR) spectroscopy, combining Faraday and Voigt geometry measurements. The application of a magnetic field ($B$) quantizes the electronic states into Landau levels (LLs). By carefully tracking the magnetic field dependence of the inter-LL transitions, we can extract important band parameters along the three principal crystal axes and reconstruct the 3D electronic structure of ZrTe$_5$ with great energy resolution. Most saliently, we demonstrate both theoretically and experimentally that the band inversion leads to a second extremum (band gap) next to the Brillouin zone center, giving rise to two distinct sets of inter-LL transitions. Our results unambiguously identify ZrTe$_5$ as a strong topological insulator (STI) at liquid helium temperature.

\begin{figure*}[t!]
\includegraphics[width=17.8cm] {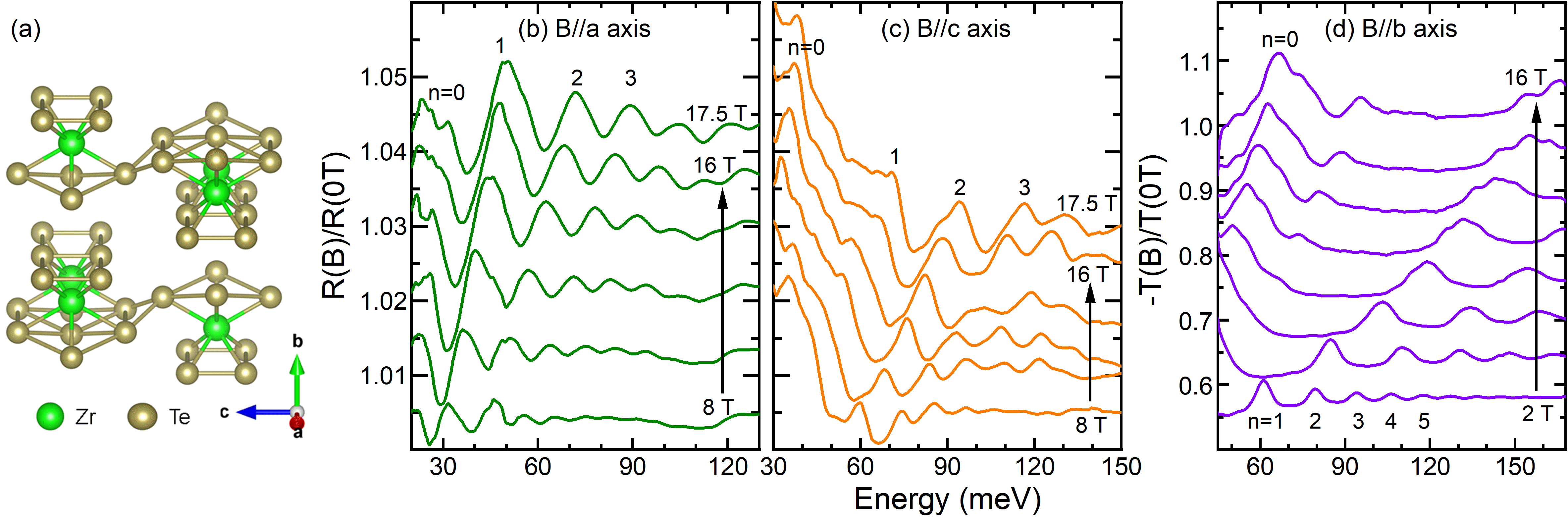}
\caption{(color online) (a) Schematic view of ZrTe$_5$ unit cell. (b-d) Normalized magneto-IR spectra of ZrTe$_5$ with the magnetic field applied along three principal crystal axes. To optimize the signal \cite{SM}, the spectra with $B \parallel$ $a$-axis (b) and $B \parallel$ $c$-axis (c) are measured in Voigt reflection, while those with $B \parallel$ $b$-axis (d) are measured in Faraday transmission. In (b-d), the interband LL transitions $L_{-n(-n-1)}\rightarrow L_{n+1(n)}$ are recognized as spectral peaks and labeled with integer $n=0, 1, 2\ ...$. All measurements are performed at liquid helium temperature ($T=4.2$ K), and the spectra are offset vertically for clarity.}
\end{figure*}

The ZrTe$_5$ single crystals studied in this work were prepared by the Te-assisted chemical vapor transport method \cite{YJ_ZT5} or molten Te flux growth \cite{jgcheng}. The orthorhombic crystal structure is illustrated in Fig. 1(a). The as-grown samples exhibit a shiny needle-like surface (typically 0.5 mm by 10 mm) in the $ac$-plane but with a thin thickness (typically 0.1 mm) along the $b$-axis. With such high aspect ratio, the commonly used Faraday geometry measurement, where light travels in the magnetic field direction, is only suitable for studying the electronic structure in the $ac$-plane, while the Voigt geometry measurement, where light travels perpendicular to $B$, is more effective for the study in the $ab$- and $bc$-planes. Specifically, to optimize the signal, we employed Faraday transmission measurements with $B \parallel$ $b$-axis and Voigt reflection measurements with $B \parallel$ $a$-axis and $B \parallel$ $c$-axis, respectively. All the measurements were performed at 4.2 K with a magnetic field up to 17.5 T. Further crystal synthesis and experimental details can be found in the Supplementary Material (SM) \cite{SM}.

Figure 1(b-d) shows the normalized magneto-IR spectra of ZrTe$_5$ with the magnetic field applied along different crystal axes. In all three cases, one can readily identify a series of peaks (or modes), which blue-shift as $B$ increases and can be attributed to specific inter-LL transitions $L_{-n(-n-1)}\rightarrow L_{n+1(n)}$ labeled by integer $n=0,1,2,...$. In Fig. 2(a-c), we summarize the magnetic field dependence of the transition energies as a function of $\sqrt{B}$ by extracting the central energy of each mode at different magnetic fields. Here, for simplicity, we focus on the central peak of each transition with the strongest optical weight and omit the weak satellite peaks from the splitting of low-lying transitions.

By comparing the three cases in Fig. 2(a-c), one can see the anisotropy in the electronic structure of ZrTe$_5$. On the one hand, the LL transition energies with $B \parallel b$-axis (Fig. 2(c)) exhibit a nearly perfect linear-in-$\sqrt{B}$ dependence, characteristic of Dirac-like dispersion. On the other hand, the transition energies with $B \parallel a$-axis (Fig. 2(a)) and $B \parallel c$-axis (Fig. 2(b)) grow much more slowly with increasing $B$ and show strong deviations from the $\sqrt B$ dependence. A closer inspection of Fig. 2(a,b) also reveals that if one linearly extrapolates (linear in $\sqrt{B}$) the $n\neq 0$ LL transitions to zero magnetic field, a negative energy intercept is obtained. This behavior is very similar to that in inverted PB semiconductors \cite{YJ_InAs}.

Quantitatively, the modes in Fig. 2(c) can be described by a massive Dirac fermion model \cite{YJ_ZT5,2DZT5_AA}, where the LL energies read
\begin{align}
\label{Dirac}
E_n=\alpha \sqrt{2 e\hbar v^2_F n B+M^2},
\end{align}
with integer $n$ being the LL index, $\alpha=\pm 1$ the band index, $e$ the electron charge, $\hbar$ the reduced Planck's constant, $v_F$ the Fermi velocity, and $M$ the Dirac mass. However, this model fails to explain the data in Fig. 2(a,b) because (1) the model predicts a positive zero-field intercept regardless of the sign of $M$; (2) for low-lying LL transitions ($n<3$), the energy ratio of two adjacent modes significantly deviates from the model prediction; and (3) for high-order LL transitions ($n>3$), the model predicts a linear-in-$\sqrt{B}$ dependence at high magnetic fields but the experimental data curve up. All these deviations from the massive Dirac fermion model suggest strong band anisotropy in ZrTe$_5$ and the necessity to extend the model to include PB contributions.

\begin{figure*}[t]
\includegraphics[width=14cm] {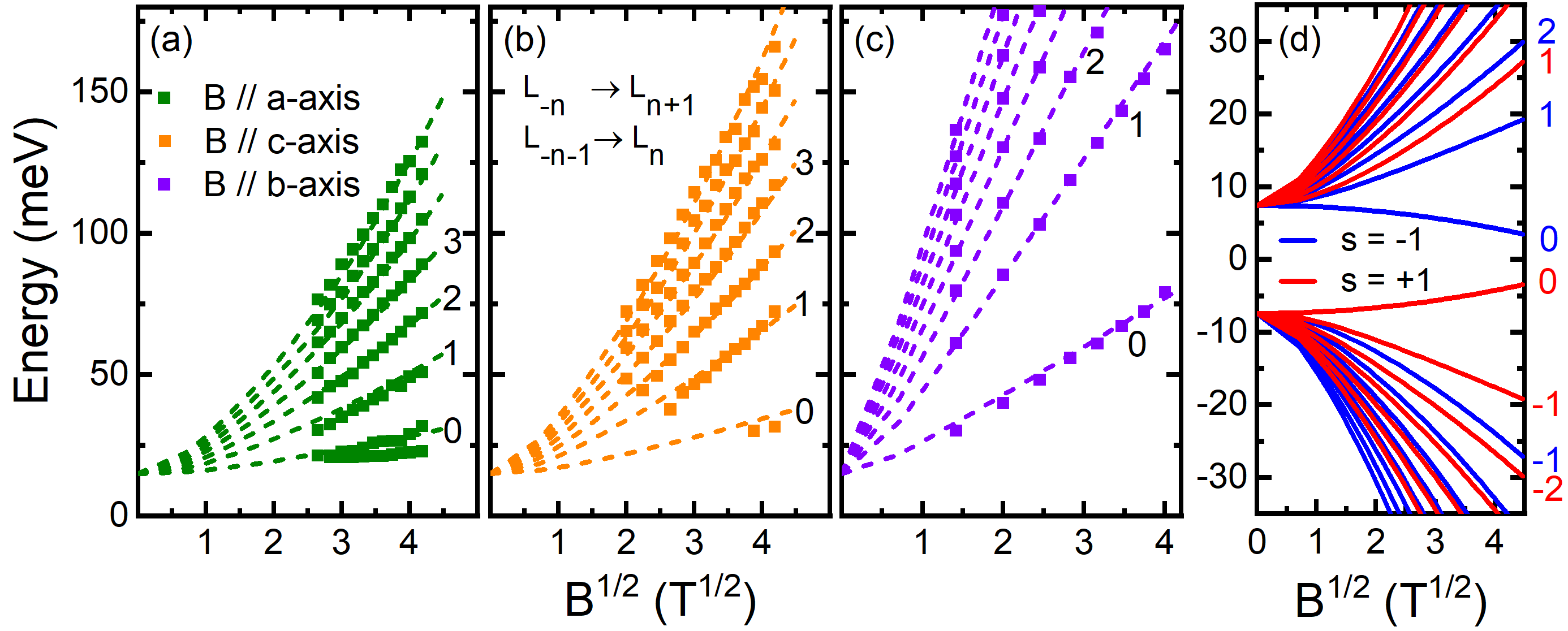}
\caption{(color online) (a-c) Magnetic field dependence of the extracted LL transition energies from Fig. 1(b-d) for $B\parallel a$-axis (a), $B\parallel c$-axis (b), and $B\parallel b$-axis (c), with the symbol size indicating the upper bound of errors in energy positions. The dash lines are best fits to the data using Eq. \eqref{LL}. When splitting occurs, the fit goes through the average energy of the two branches. The interband LL transitions $L_{-n(-n-1)}\rightarrow L_{n+1(n)}$ are labeled by integer $n=0,1,2,...$, consistent with that in Fig. 1(b-d). (d) Representative Landau fan diagram for the case of $B\parallel a$-axis. The red and blue lines correspond to the $s=\pm 1$ LLs, respectively.}
\end{figure*}

Next, we show that all our data can be well explained by a $\mathbf{k \cdot p}$ model that accounts for the symmetry at $\Gamma$ point, the PB contributions, and the material anisotropy. The effective Hamiltonian \cite{SM} reads
\begin{align*}
H(\mathbf{k})=
\begin{pmatrix}
    &L(\mathbf{k}) &0 &Ak_- &A_zk_z \\
    &0 &L(\mathbf{k}) &A_zk_z &-Ak_+ \\
    &Ak_+ &A_zk_z &-L(\mathbf{k}) &0 \\
    &A_zk_z &-Ak_- &0 &-L(\mathbf{k})
\end{pmatrix},
\end{align*}
where $L(\mathbf{k})=M-(\mathscr{B}_x k_x^2+\mathscr{B}_y k_y^2+\mathscr{B}_z k_z^2)$, $Ak_\pm=\hbar (v_{Fx} k_x \pm i v_{Fy} k_y)$, and $A_z=\hbar v_{Fz}$. The $x$-, $y$-, and $z$-directions correspond to the $a$-, $c$-, and $b$-axes of ZrTe$_5$, respectively. The electronic structure is then determined by a set of material parameters: (1) LB component $\mathbf{v_F}=(v_{Fx},v_{Fy},v_{Fz})$; (2) PB component $\mathbf{\mathscr{B}}=(\mathscr{B}_x,\mathscr{B}_y,\mathscr{B}_z)$, which is also called the band inversion parameter; and (3) Dirac mass $M$. Both $v_F$ and $\mathscr{B}$ carry anisotropy. The resulting LL spectrum of ZrTe$_5$ reads at $k_z=0$
\begin{align}
\label{LL}
\begin{split}
&E^{s}_{n=0}=s[M-\bar{\mathscr{B}}k_B^2], \\
&E^s_{n\neq0}=-s\bar{\mathscr{B}}k_B^2+\alpha \sqrt{2 \hbar^2 \bar{v}_F^2 n k_B^2+(M-M_B)^2},
\end{split}
\end{align}
where $s=\pm 1$, $k_B=\sqrt{eB/\hbar}$ is the inverse magnetic length, and $M_B=2\bar{\mathscr{B}} n k_B^2$ is the field induced gap. Representative Landau fan diagram for the case of $B\parallel a$-axis is shown in Fig. 2(d). This model is an extension of the effective TI model \cite{BHZ_Zhang1,BHZ_Zhang2}, but due to the band anisotropy, parameters $\bar{\mathscr{B}}$ and $\bar{v}_F$ now represent the geometric average of their values in the plane perpendicular to the magnetic field \cite{SM}. That is, for $B\parallel b$, $\bar{\mathscr{B}}=\sqrt{\mathscr{B}_a \mathscr{B}_c}$ and $\bar{v}_F=\sqrt{v_{Fa}v_{Fc}}$. Note that there is a sign freedom in Eq. \eqref{LL} since simultaneously reversing the signs of $\bar{\mathscr{B}}$ and $M$ will not affect the results. In this work, we fix the sign of $M$ to be positive and allow the sign of $\bar{\mathscr{B}}$ to vary. A positive (negative) $\bar{\mathscr{B}}$ represents an inverted (normal) band, respectively.

With the above model, one can produce excellent fits to the experimental data in all configurations, as shown in Fig. 2(a-c). Here, we only consider the electric dipole transitions, $\Delta n=\pm 1$ and $\Delta s=0$, while leaving the discussion of possible $\Delta n=0$ and $\Delta s=\pm 2$ transitions to SM \cite{SM}. Our fitting results of $B\parallel a$-axis and $B\parallel c$-axis clearly show that the $M_B$ term breaks the $\sqrt{B}$ energy dependence in Eq. \eqref{LL} via introducing a linear-in-$B$ mass term (PB contribution). This PB component is comparable with the LB component, suggesting a relatively small $\bar{v}_F$ and/or a relatively large $\bar{\mathscr{B}}$ in these two configurations. When $B \parallel b$-axis, however, the LB component dominates the LL transition energies due to the large $\bar{v}_F$. In this case, the PB contribution is relatively small, rendering the determination of the $\bar{\mathscr{B}}$ value less accurate in this direction \cite{note1}. More importantly, we note that the sign of $\bar{\mathscr{B}}$ cannot be solely determined from the data shown in Figs. 1 and 2, as the most influential PB contribution in Eq. \eqref{LL}, $(M-M_B)^2 \approx M_B^2$ for a small $M$, insensitive to the sign of $\bar{\mathscr{B}}$.

Fortunately, we find an intriguing scenario that can help circumvent the above situation and resolve the topological phase of ZrTe$_5$. This is shown in Fig. 3(a,b), where we plot the band dispersion along the $b$-axis with different $v_{Fb}$ values for an inverted ($\bar{\mathscr{B}}>0$) and normal ($\bar{\mathscr{B}}<0$) band, respectively. For the inverted band (Fig. 3(a)), the electronic structures exhibit a local extremum not only at $\Gamma$ point but also at a non-zero $k_b$-vector (denoted by $\zeta$ point) when $v_{Fb}$ is sufficiently small. As $v_{Fb}$ increases, the extremum at $\zeta$ point gradually disappears. On the contrary, such extremum at $\zeta$ point never occurs in the normal band case (Fig. 3(b)) regardless of the magnitude of $v_{Fb}$. From our experimental data shown in Figs. 1 and 2, one can extract the $v_F$ values along all three crystal axes with $v_{Fb}$ as small as $\sim 0.5 \times 10^5$ m/s, close to the violet lines in Fig. 3(a,b). Therefore, the presence of a second extremum at $\zeta$ point signifies band inversion in ZrTe$_5$ and provide a smoking gun evidence for the STI phase. Practically, this direct approach does not require any quantitative analysis, but solid proof of the second extremum.

\begin{figure}[t]
\includegraphics[width=8.5cm]{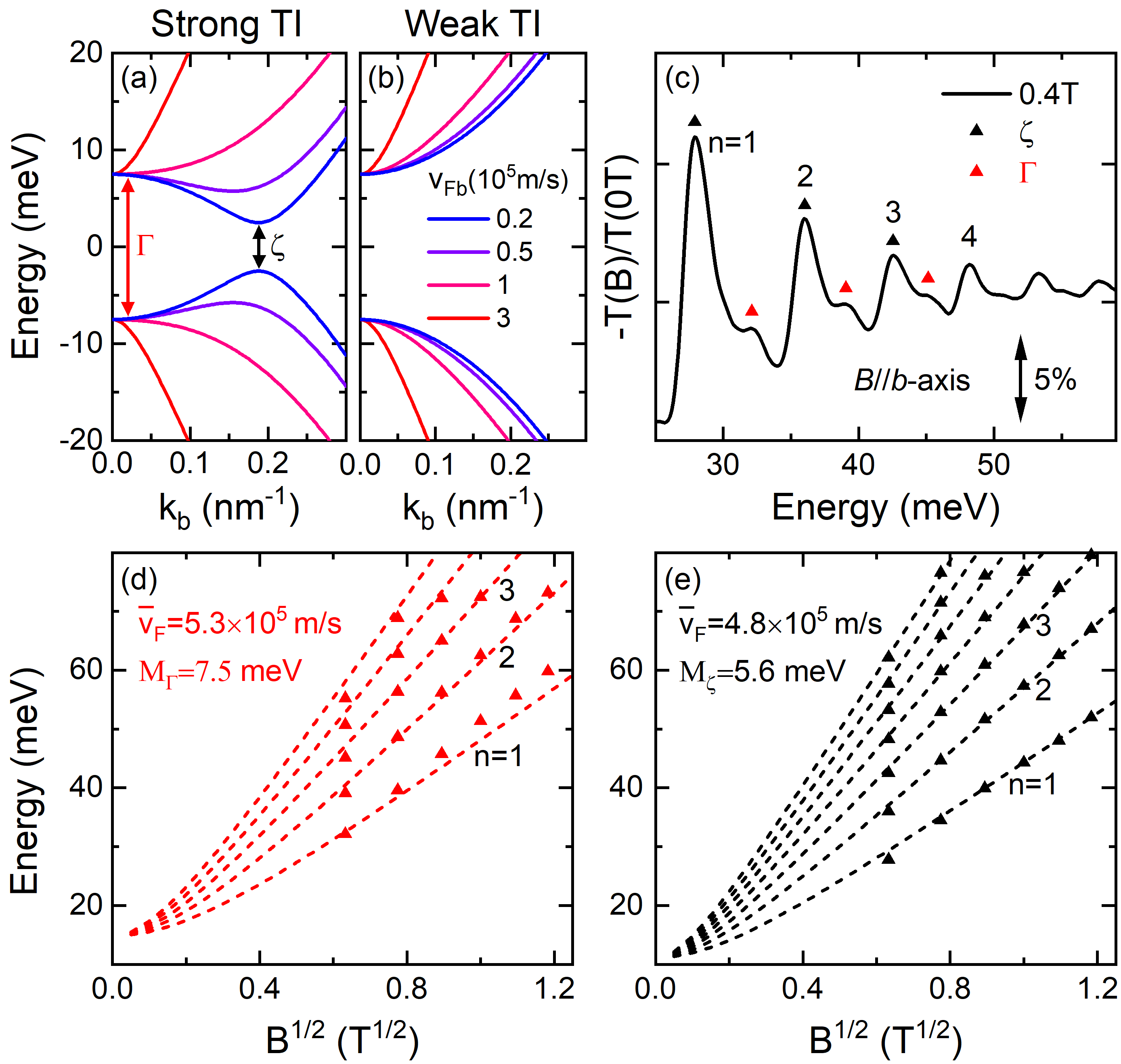}
\caption{(color online) (a,b) Zero-field band structures of ZrTe$_5$, calculated in the STI (a) and WTI (b) phases with different $v_{Fb}$. The PB component is kept the same in the calculation but with a positive (negative) sign for the STI (WTI) phase. For STI, two band extrema occur at $\Gamma$ (red arrow) and $\zeta$ (black arrow) points when $v_{Fb}$ is sufficiently small. (c) Normalized magneto-transmission spectrum, $-T(B)/T(0\text{T})$, of ZrTe$_5$ measured at $B=0.4$ T and $B \parallel b$-axis. The up-triangles label the energy positions of the LL transitions from $\zeta$ (black) and $\Gamma$ (red) points. (d,e) Best fits to the magnetic field dependence of the LL transitions from $\Gamma$ (d) and $\zeta$ (e) points using the simple massive Dirac fermion model of Eq. (\ref{Dirac}). The symbol size indicates the upper bound of errors in energy positions.}
\end{figure}

Since each local extremum in electronic structure carries a large density of states \cite{note1_5}, it can host a set of LL transitions under a magnetic field. This is indeed the case observed in our experiment when the magnetic field is applied in the slow $v_{Fb}$ direction (when $B \parallel b$-axis), as shown in Fig. 3(c). Here, due to the high quality of our ZrTe$_5$ samples, clear interband LL transitions ($n=1,2,3,...$) occur in a very low field and unambiguously exhibit a doublet structure (marked by black and red up-triangles) before the field-induced linewidth broadening takes place. The energy splitting of the doublet at such a low magnetic field cannot originate from field-induced effects such as $g$-factor or band asymmetry \cite{note2}. Linear extrapolation of the magnetic field dependence of the doublet reveals two different energy intercepts at zero field, suggesting that they belong to two distinct sets of LL transitions presumably from the $\zeta$ (black) and $\Gamma$ (red) points, respectively. With increasing $B$, the LB component becomes dominant, and the energy splitting caused by the mass difference between the $\Gamma$ and $\zeta$ points diminishes. Therefore, our results strongly suggest that the ZrTe$_5$ band is inverted, and it is in an STI phase at $T=4.2$ K.

For more quantitative analysis, one can simply fit each set of the LL transitions with the massive Dirac fermion model of Eq. (\ref{Dirac}), as a non-zero $k_b$ effectively renormalizes the Dirac mass in our anisotropic model \cite{SM}. Figure 3(d,e) shows best fits to the data, where the extracted energy gaps are $\Delta_{\Gamma}=2M_{\Gamma}=15$ meV and $\Delta_{\zeta}=2M_{\zeta}=11.2$ meV, respectively. The extracted $\bar{v}_F$ and $M_{\zeta}$ values in Fig. 3(e) are also consistent with the fitting result of Fig. 2(c). With this information and using the zero-field $\mathbf{k \cdot p}$ model, one can further deduce $\mathscr{B}_b=0.21$ eV$\cdot$nm$^2$ \cite{SM}. By combining with the fitting results of Fig. 2(a,b) along the $a$- and $c$-axes, the anisotropy of ${\mathscr{B}}$ is now fully resolved with improved accuracy. Similarly, the anisotropy of $v_F$ at $\Gamma$ point can be deduced using the extracted $\bar{v}_F$ values from Fig. 3(d) and Fig. 2(a,b).

\begin{table}[t!]
\caption{Extracted band parameters (at $\Gamma$ point) along the three principal crystal axes of ZrTe$_5$ using the anisotropic $\mathbf{k \cdot p}$ model.}
\begin{tabular}{|C{3cm}| C{1.5 cm} | C{1.5 cm} | C{1.5 cm} |}
\hline 
 ZrTe$_5$ @ $\Gamma$ point & $k \parallel a$ & $k \parallel b$ & $k \parallel c$\\
\hline
$v_F$ (10$^5$m/s)&6.85 &0.50 &4.10 \\
$\mathscr{B}_i$(eV$\cdot$nm$^2$) &0.12 &0.21 &0.08\\
$M_i$(meV)& 7.5 & 7.5 & 7.5\\
\hline
\end{tabular}
\end{table}

In Table I, we summarize the band parameters of ZrTe$_5$ along the three principal crystal axes. Owing to the small $v_{Fb}$, the low-energy electronic structure along the $b$-axis is dominated by the PB component, leading to a much weaker dispersion than those along the LB dominated $a$- and $c$-directions. Our findings are consistent with recent transport and IR experiments \cite{Tp_WP,Tp_FXX,Tp_MLT,JY_PNAS,2DZT5_AA}.

In conclusion, we have performed a magneto-IR spectroscopy study of the band anisotropy in ZrTe$_5$. We find that the LB dispersion (characterized by $v_F$) along the $b$-axis is about one order of magnitude smaller than those along the $a$- and $c$-axes. When $v_F$ is small, the PB component can strongly modify the band dispersion, and the presence of a second extremum in $b$-direction indicates band inversion. Incorporating prior results of first-principles calculations, we further identify ZrTe$_5$ as an STI at low temperatures. Our work provides an experimental approach to directly infer the topological phase of anisotropic materials from their bulk band structure.

This work was primarily supported by the DOE (Grant No. DE-FG02-07ER46451), while crystal growth at GT and UT were supported by Grant No. DE-SC0018660 and DE-SC0020254. Crystal characterization was performed in part at the GT Institute for Electronics and Nanotechnology, a member of the National Nanotechnology Coordinated Infrastructure, which is supported by the NSF (Grant No. ECCS-1542174). IR measurements were performed at NHMFL, which is supported by the NSF Cooperative Agreement No. DMR-1644779 and the State of Florida. Y.J. acknowledges support from the NHMFL Jack Crow Postdoctoral Fellowship, and Z.J. acknowledges the NHMFL Visiting Scientist Program. J.W. acknowledges support from the China Scholarship Council. Work at Sandia is supported by a Laboratory Directed Research and Development project and a user project at the Center for Integrated Nanotechnologies, an Office of Science User Facility operated for the DOE Office of Science. Sandia National Laboratories is a multimission laboratory managed and operated by National Technology and Engineering Solutions of Sandia, LLC., a wholly owned subsidiary of Honeywell International, Inc., for the U.S. DOE's National Nuclear Security Administration under contract DE-NA-0003525. This paper describes objective technical results and analysis. Any subjective views or opinions that might be expressed in the paper do not necessarily represent the views of the U.S. DOE or the U.S. Government.


\begin{thebibliography}{99}
\bibitem{note1}Note that the Dirac mass obtained from fitting the data in Fig. 2(c) is slightly smaller than that from Fig. 2(a,b). This is because the band gap along the $b$-axis is not at $\Gamma$ point, as discussed in the context of Fig. 3.
\bibitem{note1_5}The $\zeta$ point assigned here is a second extremum of the same band as the $\Gamma$ point. Due to its proximity to the $\Gamma$ point, $\zeta$ point is different from the $Z$ point reported in Refs. \cite{Arpes1_MG,Arpes2_XJZ}, which belongs to another band across the Fermi energy.
\bibitem{note2}Based on our prior results \cite{YJ_ZT5}, we estimate the splitting of the $n=1$ LL transition, due to $g$-factor or band asymmetry, $\sim 0.4$ meV, one order of magnitude smaller than the 4 meV splitting observed in Fig. 4(c).
\end{thebibliography}
\end{document}


\title{Supplementary Material: Unraveling the Topological Phase of ZrTe$_5$ via Magneto-infrared Spectroscopy}
\author{Y. Jiang}
\affiliation{National High Magnetic Field Laboratory, Tallahassee, Florida 32310, USA}
\author{J. Wang}
\affiliation{School of Physics, Georgia Institute of Technology, Atlanta, Georgia 30332, USA}
\affiliation{State Key Laboratory for Artificial Microstructure and Mesoscopic Physics, Peking University, Beijing 100871, China}
\author{T. Zhao}
\affiliation{School of Physics, Georgia Institute of Technology, Atlanta, Georgia 30332, USA}
\author{Z. L. Dun}
\affiliation{School of Physics, Georgia Institute of Technology, Atlanta, Georgia 30332, USA}
\author{Q. Huang}
\affiliation{Department of Physics and Astronomy, University of Tennessee, Knoxville, Tennessee 37996, USA}
\author{X. S. Wu}
\affiliation{State Key Laboratory for Artificial Microstructure and Mesoscopic Physics, Peking University, Beijing 100871, China}
\author{M. Mourigal}
\affiliation{School of Physics, Georgia Institute of Technology, Atlanta, Georgia 30332, USA}
\author{H. D. Zhou}
\affiliation{Department of Physics and Astronomy, University of Tennessee, Knoxville, Tennessee 37996, USA}
\author{W. Pan}
\affiliation{Quantum and Electronic Materials Department, Sandia National Laboratories, Livermore, California 94551, USA}
\author{M. Ozerov}
\affiliation{National High Magnetic Field Laboratory, Tallahassee, Florida 32310, USA}
\author{D. Smirnov}
\affiliation{National High Magnetic Field Laboratory, Tallahassee, Florida 32310, USA}
\author{Z. Jiang}
\email{zhigang.jiang@physics.gatech.edu}
\affiliation{School of Physics, Georgia Institute of Technology, Atlanta, Georgia 30332, USA}
\date{\today}

\maketitle

\section{Sample growth and experiment setup}
In this work, the single-crystal ZrTe$_5$ samples were synthesized by both the Te-assisted chemical vapor transport (CVT) method \cite{YJ_ZT5} and the molten Te flux growth \cite{jgcheng}. For the CVT growth, we first prepared polycrystalline ZrTe$_5$ by reacting appropriate ratio of Zr ($\geq 99.9\%$) and Te ($\geq 99.9999\%$) in an evacuated quartz tube at 450 $^{\circ}$C for one week. Next, 2 g of polycrystalline ZrTe$_5$ and transport agent (100 mg 99.9999\% Te) were sealed in a quartz tube and placed horizontally in a tube furnace. The sample (source) was placed at the center of the furnace and was heated up to 530 $^{\circ}$C at a rate of 60 $^{\circ}$C per hour. The growth zone (sink) was 12 cm away, and its temperature was measured to be 450 $^{\circ}$C.

In the flux growth, elemental Zr ($\geq 99.9\%$) and Te ($\geq 99.9999\%$) were directly mixed in the molar ratio of 1:49 and sealed under vacuum in a quartz tube. The mixture was heated to 1000 $^{\circ}$C at a rate of 100 $^{\circ}$C per hour, and held at this temperature for 12 hours. Then, we cooled down the mixture in two steps: first to 650 $^{\circ}$C at a rate of 100 $^{\circ}$C per hour; and next to 460 $^{\circ}$C at a rate of 3 $^{\circ}$C per hour. Finally, the crystals were separated from the flux by centrifuging at 460 $^{\circ}$C.

Broad-band magneto-infrared (magneto-IR) spectroscopy measurements were performed using a Bruker 80v Fourier-transform IR spectrometer. The unpolarized IR light from a globar source was delivered to the sample through evacuated light pipes. Lights were focused onto the sample and collected into a Si bolometer detector using parabolic focusing cones. The sample was placed at the center of a $B=17.5$ T superconducting magnet in a helium exchange gas environment. During the measurement, IR spectra were taken at selected magnetic fields and normalized to the zero-field spectrum to show the magnetic field induced change.

Figure S1 shows the schematics of two experiment configurations. In Faraday configuration (where the light propagation direction is parallel to the magnetic field), transmission measurements were performed with $B \parallel b$-axis. The ZrTe$_5$ sample was prepared by repeatedly exfoliating the material with an IR-transparent Scotch tape. The resulting ZrTe$_5$ flakes have an average thickness of $\sim$1 $\mu$m. Although transmission measurement is also preferred in Voigt configuration (where the light propagation direction is perpendicular to the magnetic field), one has to align the thin ZrTe$_5$ flakes in the same orientation while maintaining a relatively uniform thickness and full coverage on the tape. This is a challenging task. Instead, we carried out reflection measurements in Voigt configuration for $B \parallel$ $a$-axis and $B \parallel$ $c$-axis, in which we only need to align fewer, larger-size ZrTe$_5$ crystals on the tape. 
\begin{figure*}[t!]
\includegraphics[width=0.7\textwidth] {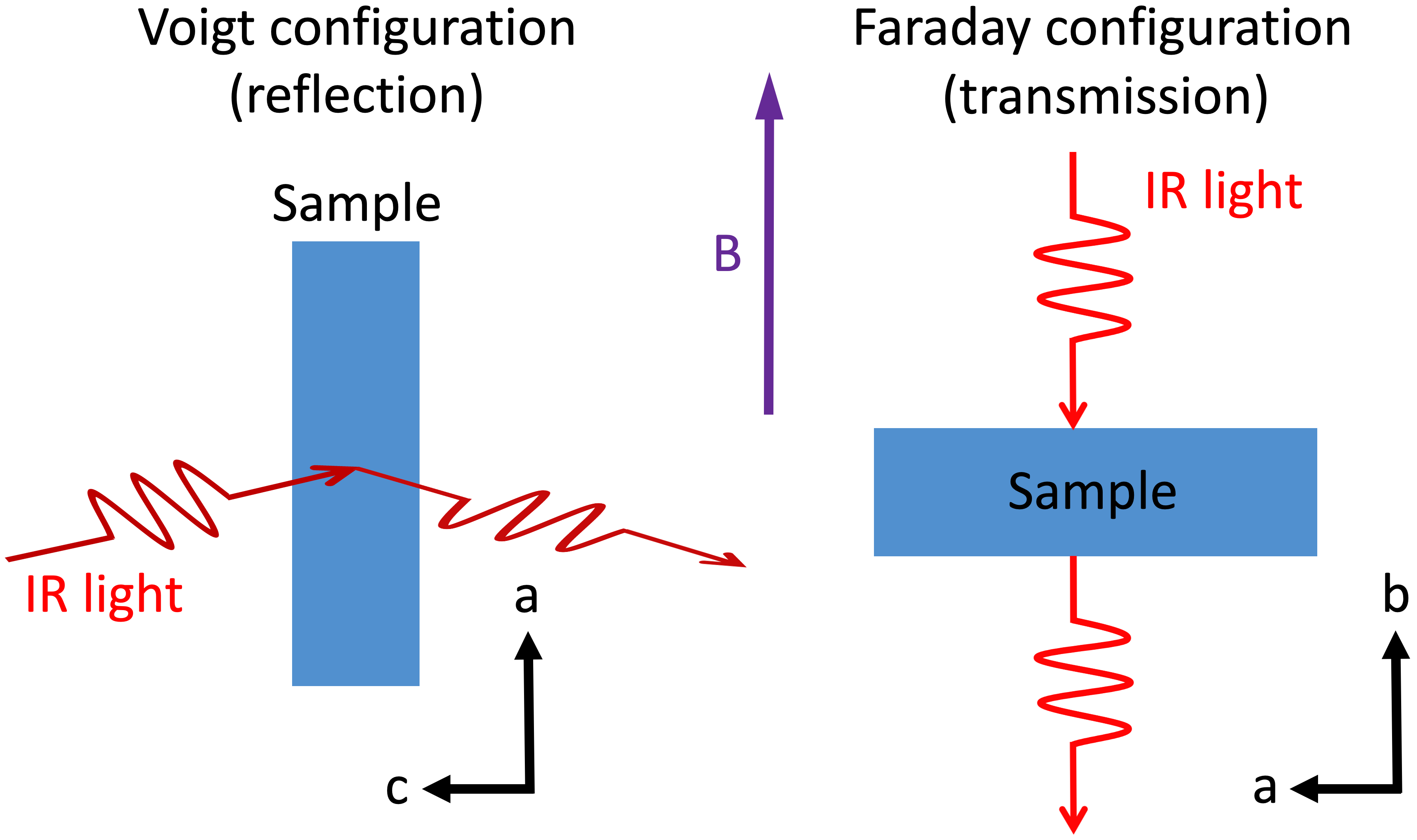}
\caption{(color online) Schematics of Voigt and Faraday measurement configurations. In the Voigt (Faraday) configuration, the incident and reflected (transmitted) IR lights travel in the $bc$-plane ($ab$-plane) while the magnetic field $B$ is applied along the $a$-axis ($b$-axis). $a$, $b$, and $c$ label the principal crystal axes of ZrTe$_5$.}
\end{figure*}

\section{Excluding contributions from trivial bands}
IR spectroscopy has proven to be an effective tool in identifying topological phases of quantum materials and studying topological phase transitions. One example relevant to this work is the study of the crossover between weak and strong topological crystalline insulator phases in Pb$_{1-x}$Sn$_x$Se \cite{PbSnSe_1}. When combined with a magnetic field, magneto-IR spectroscopy becomes one of the most accurate methods in determining a small bandgap, on the order of 10 meV, which is crucial for the reported study herein. Additionally, via following the magnetic field dependence of Landau level (LL) transitions, one can separate the optical responses of the Dirac-like non-trivial band from those of trivial bands in different spectral regions. In the case of ZrTe$_5$, the Dirac-like band is expected to be near the $\Gamma$ point and exhibiting a fast magnetic field dispersion. Other trivial parabolic bands near Fermi energy are expected to show a cyclotron-resonance-like LL transition, which features a slow and linear-in-$B$ dependence due to the relatively large effective mass of the carriers in the bands. Therefore, by tracking the fast blueshifted LL transitions in the far- and mid-IR region with an increasing magnetic field, one can investigate (solely) the properties of the non-trivial band near $\Gamma$ point.

\section{Anisotropic $\mathbf{k \cdot p}$ model}
In this section, we introduce the $\mathbf{k \cdot p}$ model describing the anisotropy of ZrTe$_5$. The anisotropic Hamiltonian was previously constructed \cite{YJ_ZT5,Refl_NLW} up to $k^2$ terms with the following bases: $\left( \ket{+,\uparrow}, \ket{+,\downarrow},\ket{-,\uparrow}, \ket{-,\downarrow}\right)$, where the orbitals and spins are denoted by $\pm$ and $\uparrow,\downarrow$, respectively. The three-dimensional (3D) anisotropic $\mathbf{k \cdot p}$ Hamiltonian for ZrTe$_5$ reads
\begin{align}
\label{3DHamil}
H(\bold{k})=&(M-\Sigma_{i}\mathscr{B}_i k^2_i) \tau^z
+\hbar(v_{Fx}k_x\tau^x\sigma^z+v_{Fy}k_y\tau^y+v_{Fz}k_z \tau^x \sigma^x).\tag{S1}
\end{align}
Here, $\mathbf{k}=(k_x,k_y,k_z)$ is the wave vector, index $i=x,y,z$ labels the three principal axes ($a$-, $c$-, and $b$-axes) of ZrTe$_5$, $\hbar$ is the reduced Planck's constant, and $\tau$ and $\sigma$ are the Pauli matrices for orbital and spin, respectively. The material parameters are the same as those described in the main text: (1) Fermi velocity $\mathbf{v_F}=(v_{Fx},v_{Fy},v_{Fz})$; (2) band inversion parameter $\mathbf{\mathscr{B}}=(\mathscr{B}_x,\mathscr{B}_y,\mathscr{B}_z)$; and (3) Dirac mass $M$. Both $v_F$ and $\mathscr{B}$ carry anisotropy.

By diagonalizing the Hamiltonian, one can arrive at the 3D zero-field band dispersion around $\Gamma$ point
\begin{align*}
E(\mathbf{k})=\alpha \sqrt{\Sigma_{i} \hbar^2v_{Fi}^2k_i^2+(M-\Sigma_{i} \mathscr{B}_i k^2_i)^2},
\end{align*}
where $\alpha=\pm1$ denotes conduction and valence bands, respectively. The dispersion along each principal axis can then be described as
\begin{align}
\label{ZF}
E=\alpha \sqrt{\hbar^2v_{F}^2k^2+(M- \mathscr{B} k^2)^2}.\tag{S2}
\end{align}

Particularly, around $\zeta$ point (0,0,$\zeta$), the dispersion in the $k_x-k_y$ plane can be written as
\begin{align*}
E(\mathbf{k})=\alpha \sqrt{\hbar^2 v_{Fx}^2 k_x^2+\hbar^2 v_{Fy}^2 k_y^2+M_{\zeta}^2}
\end{align*}
where $M_{\zeta}^2=\hbar^2 v_{Fz}^2k_{\zeta}^2+(M-\mathscr{B}_z k^2_{\zeta}-\mathscr{B}_x k^2_x-\mathscr{B}_y k^2_y)^2 $. If $ k_x \ll \sqrt{M/\mathscr{B}_x} $ and $ k_y \ll \sqrt{M/\mathscr{B}_y} $, then $M_{\zeta}^2 \approx \hbar^2 v_{Fz}^2k_{\zeta}^2+(M-\mathscr{B}_z k^2_{\zeta})^2$. From Table I of the main text, one can estimate $\sqrt{M/\mathscr{B}_x} \sim \sqrt{M/\mathscr{B}_y} \sim$ 0.28 nm$^{-1}$, which is equivalent to a magnetic field of 52 T in terms of magnetic length. Therefore, the above approximation is justified in our low-field fitting, and one can model the charge carriers near $\zeta$ point as massive Dirac fermions with a renormalized Dirac mass $M_{\zeta}$.

\section{Landau levels in anisotropic $\mathbf{k \cdot p}$ model}
We now consider the magnetic field effects. We first consider the $B \parallel b$-axis case, and for convenience, we make the following scaling:
\begin{align}
\label{change}
\mathscr{B}_i=\mathscr{B'}_i\hbar^2v_{Fi}^2.\tag{S3}
\end{align}
To produce analytic solution, we need to make an approximation by replacing the in-plane anisotropy $\mathscr{B'}_x, \mathscr{B'}_y$ with an effective average $\mathscr{\bar{B}'}$, which we will further discuss in the next section.

We consider the extremum point $k_z=0$ (or $k_b=0$) as its large density of states provides dominant contributions in the magneto-optical response. The Hamiltonian can be transformed into two decoupled diagonal blocks by a unitary transformation 
\begin{align*}
H'=U^{\dagger}HU, \qquad \text{with}\qquad U= 
\begin{pmatrix}
&1 &0 &0 &0\\
&0 &0 &1 &0\\
&0 &1 &0 &0\\
&0 &0 &0 &1\\
\end{pmatrix},  
\end{align*}
which essentially rearranges the basis function into $\left( \ket{+,\uparrow}, \ket{-,\uparrow},\ket{+,\downarrow}, \ket{-,\downarrow}\right)$.
The transformed Hamiltonian now reads $
H'=\begin{pmatrix}
&H_1 &0\\
&0 &H_2\\
\end{pmatrix}$, where
\begin{align*}
H_1=
\begin{pmatrix}
    &M-\mathscr{\bar{B}'}\hbar^2(v_{Fx}^2k_x^2+v_{Fy}^2k_y^2)   &\hbar v_{Fx} k_x+i\hbar v_{Fy}k_y \\
    &\hbar v_{Fx}k_x-i\hbar v_{Fy}k_y       &-M+\mathscr{\bar{B}'}\hbar^2(v_{Fx}^2k_x^2+v_{Fy}^2k_y^2)
\end{pmatrix}.    
\end{align*}

To obtain the LL spectrum, we perform the standard Peierls substitution using the Landau-gauge vector potential $\bold{A}=(-By,0,0)$. By defining
\begin{align}
\label{ladder1}
  \Delta=\sqrt{2e\hbar v_{Fx} v_{Fy} B} \quad \text{and} \quad  \hat{b}= \hbar (v_{Fx}\hat{k}_x-iv_{Fy}\hat{k}_y)/ \Delta, \tag{S4}
\end{align}
with $e$ being the electron charge, we get
\begin{align}
\label{ladder2}
[\hat{b},\hat{b}^{\dagger}]=1 \quad \text{and} \quad \hbar^2 (v_{Fx}^2k^2_x+v_{Fy}^2k^2_y)=(\hat{b}^{\dagger}\hat{b}+\frac{1}{2})\Delta^2. \tag{S5}
\end{align}
Next, we implement the ansatz that the trial wavefunction for $H_1$ is $\psi_{\uparrow}=(c_1 \phi_n, c_2\phi_{n-1})$, where $\phi_n$ is the eigenwavefunction of Harmonic oscillator, $n$ denotes the LL index, and $c_1$ and $c_2$ are the coefficients to be determined. Now, $H_1$ in a magnetic field becomes
\[
H_1=
\begin{pmatrix}
    &M-\mathscr{\bar{B}'}(n+\frac{1}{2})\Delta^2   &\Delta \sqrt{n} \\
    &\Delta \sqrt{n}                                        &-M+\mathscr{\bar{B}'}(n-\frac{1}{2})\Delta^2
\end{pmatrix}.
\]
The LL energies and the coefficients $c_1$ and $c_2$ can be calculated from solving the eigenvalue problem of this algebraic Hamiltonian. $H_2$ can be calculated in a similar fashion. Altogether, we have the following LL energies,
\begin{align}
\label{kpLL}
\begin{split}
&E_0^{\uparrow}=M-\frac{\mathscr{\bar{B}'}}{2}\Delta^2,\qquad E_0^{\downarrow}=-M+\frac{\mathscr{\bar{B}'}}{2}\Delta^2,\\
& E^s_{n\neq 0,\alpha}=-\frac{s\mathscr{\bar{B}'}}{2}\Delta^2 +\alpha \sqrt{\Delta^2 n+\left(M-n\mathscr{\bar{B}'}\Delta^2\right)^2},
\end{split} \tag{S6}
\end{align}
where $s=\pm 1$. Comparing to the LLs in the main text, one can immediately see that
$\bar{v}_F=\sqrt{v_{Fa} v_{Fc}}$ and $\mathscr{\bar{B}}=\mathscr{\bar{B}'} \hbar^2 \bar{v}_F^2$. 

A similar procedure can be applied to the $B \parallel a$-axis and $B \parallel c$-axis cases, and the resulting LLs have similar expressions. There is only one subtlety: when $B \parallel c$-axis, $s$ no longer denotes the spin.  To see this, one can make a unitary transformation $U=e^{i\sigma^x \frac{\pi}{4}}$ to rotate the spin directions, leading to $\sigma^x \rightarrow \sigma^x,\sigma^y \rightarrow -\sigma^z,\sigma^z \rightarrow \sigma^y$. The new Hamiltonian reads
\begin{align*}
H(\bold{k})=&(M-\Sigma_{i}\mathscr{B}_i k^2_i) \tau^z
+\hbar(v_{Fx}k_x\tau^x\sigma^y+v_{Fy}k_y\tau^y+v_{Fz}k_z \tau^x \sigma^x).
\end{align*}
At $k_y=0$ (or $k_c=0$), it is again possible to rotate the Hamiltonian into a diagonal block form  $H'=\begin{pmatrix}
&H_{1} &0\\
&0 &H_{2}\\
\end{pmatrix}$ with another unitary transformation
$\begin{pmatrix}
&1 &0 &0 &0\\
&0 &0 &1 &0\\
&0 &0 &0 &1\\
&0 &1 &0 &0\\
\end{pmatrix}$. This is equivalent to transform the basis to $\left( \ket{+,\uparrow}, \ket{-,\downarrow},\ket{+,\downarrow}, \ket{-,\uparrow}\right)$. As one can see, the upper and lower blocks are no longer composed of spin-polarized states. Therefore, the spin is no longer a good quantum number to differentiate the LLs within the same index.

\section{Extracting the anisotropic band parameters}
In this section, we discuss how we extract the anisotropic band parameters from the experiment. To avoid confusion, we note that the barred variables are the fitting parameters from the data with their subscripts indicating the magnetic field directions, while the unbarred variables are the band parameters (summarized in Table I of the main text). 

The extraction of $v_F$ is straightforward. Since the relation between the fitting parameters and the Fermi velocity are $\bar{v}_{Fi}=\sqrt{v_{Fj} v_{Fk}}$, where $i,j,k=a,b,c$, one can translate them into $v_{Fi}=(\bar{v}_{Fj} \bar{v}_{Fk})/ \bar{v}_{Fi}$.

The extraction of $\mathscr{B}$ requires some additional information. First, although one can directly extract $\mathscr{\bar{B}}_a$ and $\mathscr{\bar{B}}_c$ from the fitting results of Fig. 2(a,b) of the main text, the extraction of $\mathscr{\bar{B}}_b$ for the $B \parallel b$-axis case from Fig. 2(c) of the main text is less accurate, because (1) the observed LL transitions are actually from the $\zeta$ point, not the $\Gamma$ point of the Brillouin zone; and (2) the large $\bar{v}_{Fb}$ in this direction makes the linear band component dominate the electronic structure. Instead, we extract ${\mathscr{B}_b}$ directly with Eq. \eqref{ZF} from the gap difference between the $\zeta$ and $\Gamma$ points.

Second, we do not know the exact relation between the fitting parameters and the band inversion parameter. To proceed, we rewrite the anisotropic parabolic band terms in the diagonal elements of the LL Hamiltonian using the relations \eqref{ladder1} and \eqref{ladder2}:
\begin{align}
\label{diag}
M-\hbar^2(\mathscr{B}'_x v_{Fx}^2k_x^2+\mathscr{B}'_y v_{Fy}^2k_y^2)=M-\frac{\mathscr{B}'_x+\mathscr{B}'_y}{2}(b^\dagger b+\frac{1}{2})\Delta^2-\frac{\mathscr{B}'_x-\mathscr{B}'_y}{4}({b^\dagger}^2+b^2)\Delta^2, \tag{S7}
\end{align}
where if $\left|\frac{\mathscr{B}'_x-\mathscr{B}'_y}{\mathscr{B}'_x+\mathscr{B}'_y}\right|\ll 1$, one can neglect the last term. This is similar to the commonly known axial approximation \cite{winkler}. But, in strongly anisotropic materials (like ZrTe$_5$), such a presumption is not guaranteed, and axial approximation may not be a valid approximation. Therefore, in this work, we proceed with a simple ``geometric average'', that is, $\mathscr{\bar{B}'}_i=\sqrt{{\mathscr{B}'_j}{\mathscr{B}'_k}}$ or equivalently 
$\mathscr{\bar{B}}_i=\sqrt{{\mathscr{B}_j}{\mathscr{B}_k}}$. This idea is similar to the Onsager reciprocal relation, where we construct an isotropic electronic structure with averaged band parameters while keeping the same area enclosed by the constant energy contour as in the anisotropic structure. This process effectively takes into account the last two terms in Eq. \eqref{diag}. In addition, the geometric average approximation approaches the axial approximation when $\mathscr{B}'_x \approx \mathscr{B}'_y$.

Lastly, with the information of $\mathscr{\bar{B}}_a$, $\mathscr{\bar{B}}_c$, ${\mathscr{B}}_b$, and the geometric average approximation, one can obtain $\mathscr{B}_a$ and $\mathscr{B}_c$ through the following relations
$\mathscr{B}_a=\bar{\mathscr{B}}^2_c/\mathscr{B}_b$ and $\mathscr{B}_c=\bar{\mathscr{B}}^2_a/\mathscr{B}_b$.
We note that the accuracy of the geometric average approximation may be affected by the warping of the band structure. Also, the $\zeta$ point may be far away from the $\Gamma$ point and thus not well captured with our $\mathbf{k \cdot p}$ model. A more accurate determination may require first-principles calculations, which is beyond the scope of our work. Still, the geometric average approximation provides an analytical relation and easy means of understanding the band anisotropy of ZrTe$_5$.

\section{Selection rules}
\begin{figure*}[b!]
\includegraphics[width=0.8\textwidth] {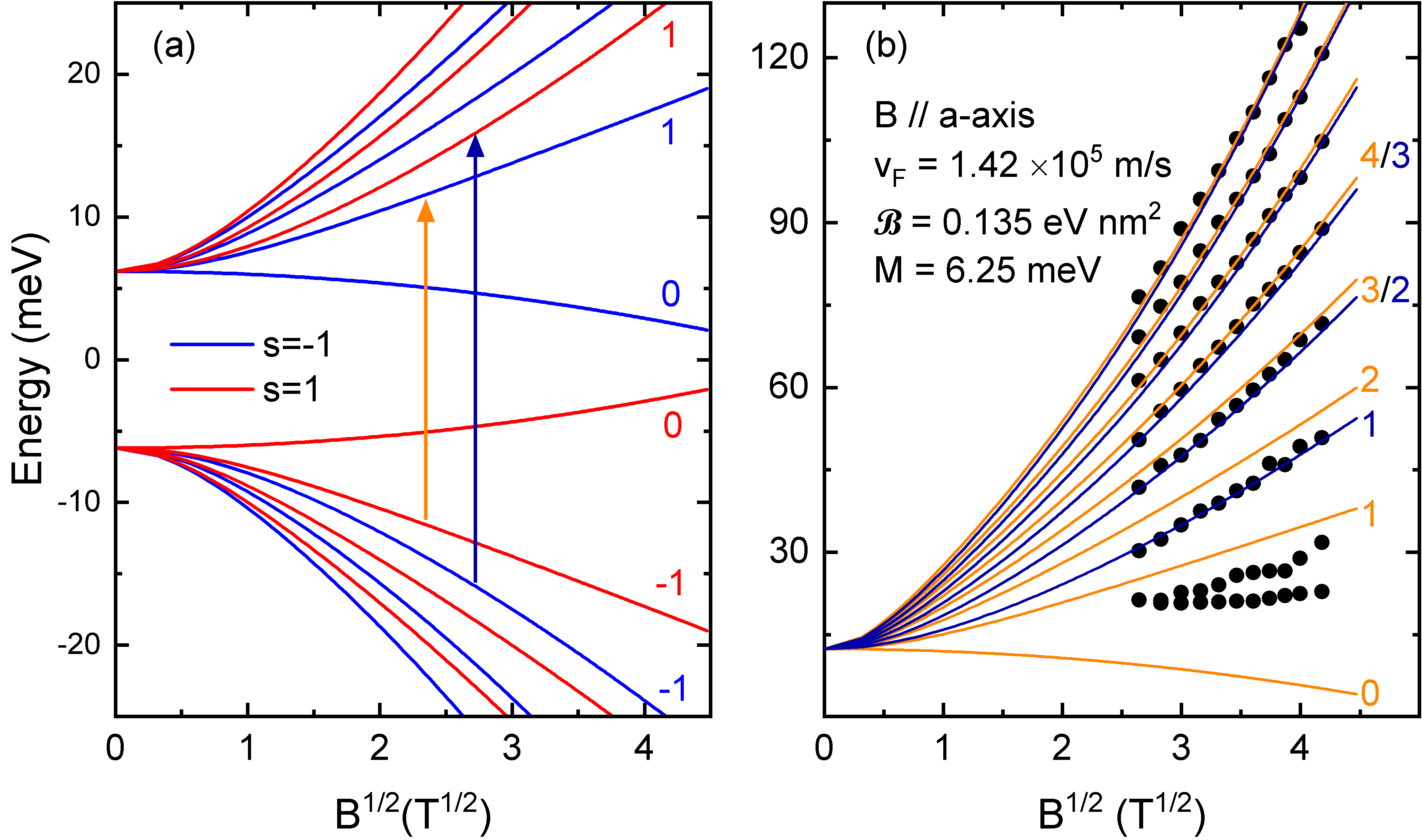}
\caption{(color online) LL transitions with $E \parallel B$ excitation. (a) Landau fan diagram for the case of $B\parallel a$-axis under $E \parallel B$ excitation. The LLs are calculated with the band parameters shown in (b) and color-coded in red ($s=+1$ LLs) and blue ($s=-1$ LLs), respectively. The dark blue (orange) arrows illustrate the allowed transitions following the $\Delta s = +2$ ($\Delta s = -2$) selection rule. (b) Best fits to the $B\parallel a$-axis data (Fig. 2(a) of the main text) with $E \parallel B$ transitions. The allowed transitions $L_{-n}\rightarrow L_n$ are labeled with LL index $n=0,1,2,3...$ and with the same color code as in (a).}
\end{figure*}
The selection rule is determined by the matrix element $\bra{I}\frac{\partial{H}}{\partial{k_i}}\ket{F}$, where $\ket{I}$ and $\ket{F}$ are the initial and final states, and $i$ denotes the electric field ($E$) polarization of the incident light. Here, we study the case of $B \parallel z$, but the conclusion also applies to the $B \parallel x,y$ cases.

Specifically, if $E \perp B$, then
$$
    \frac{\partial{H}}{\partial{k_x}}=\hbar v_{Fx}\tau^x\otimes \sigma^z,\qquad
    \frac{\partial{H}}{\partial{k_y}}=-\hbar v_{Fy}\tau^y,
    $$ 
which couples the bases with the same $s$. The $k^2$ terms in the Hamiltonian do not contribute as we are close to $\Gamma$ point. Therefore, the selection rule for $E \perp B$ reads $\Delta s=0$ and $\Delta n=\pm 1$, which represents the conventional electric dipole transitions.

If $E \parallel B$, then
$$
    \frac{\partial{H}}{\partial{k_z}}=\hbar v_{Fz} \tau^x \otimes \sigma^x,
$$
which couples the bases with opposite $s$.  Therefore, the selection rule for $E \parallel B$ is $\Delta s=\pm 2$ and $\Delta n=0$. 

In Fig. S2, we show the possible LL transitions with $E \parallel B$ excitation and their best fits to the experimental data in the $B\parallel a$-axis case (i.e., the data in Fig. 2(a) of the main text). Here, although the fits to high-order LL transitions work well, those to low-lying LL transitions deviate from the data significantly. Therefore, we exclude these $\Delta n = 0$, $\Delta s = \pm 2$ transitions when analyzing the data in the main text.

\section{Additional data from the low field measurements}
\begin{figure*}[b!]
\includegraphics[width=0.85\textwidth] {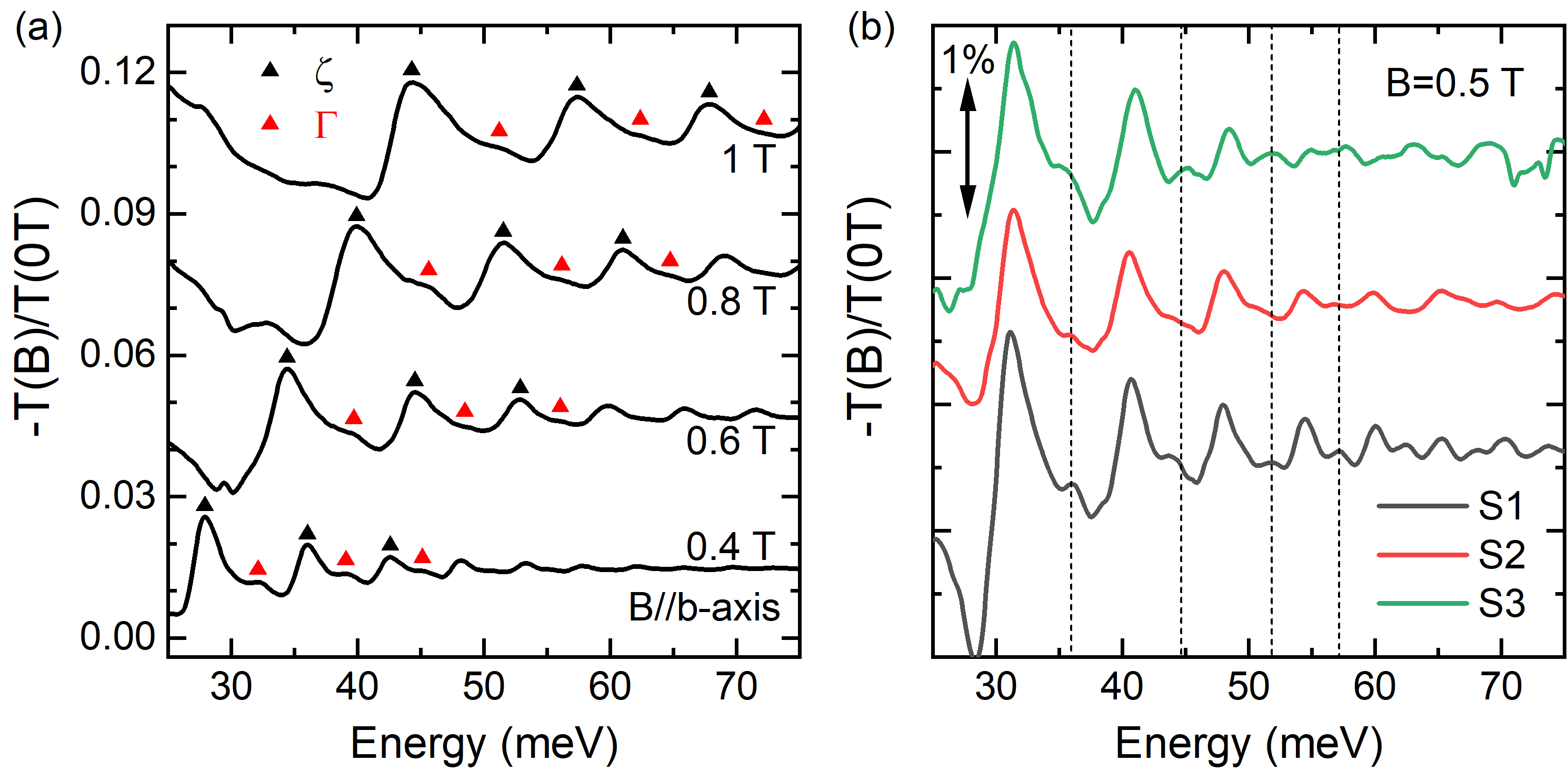}
\caption{(color online) (a) Low-field normalized magneto-transmission spectra, $-T(B)/T(0\text{T})$, of ZrTe$_5$ measured with $B \parallel b$-axis. The up-triangles label the energy positions of the LL transitions from $\zeta$ (black) and $\Gamma$ (red) points, as described in the context of Fig. 3 of the main text. (b) Magneto-transmission spectra across three different samples S1, S2, and S3 at $B=0.5$ T. The S1 sample is used in the main text. The dash lines indicate the positions of the $\Gamma$ point transition energies.}
\end{figure*}
Figure S3(a) shows the normalized magneto-transmission spectra at low magnetic fields complementary to Fig. 3(c) of the main text. The doublet structure at low fields can be consistently reproduced in several samples, as shown in Fig. S3(b).

\section{Optical weight between $\zeta$ and $\Gamma$ points}
\begin{figure*}[b!]
\includegraphics[width=\textwidth] {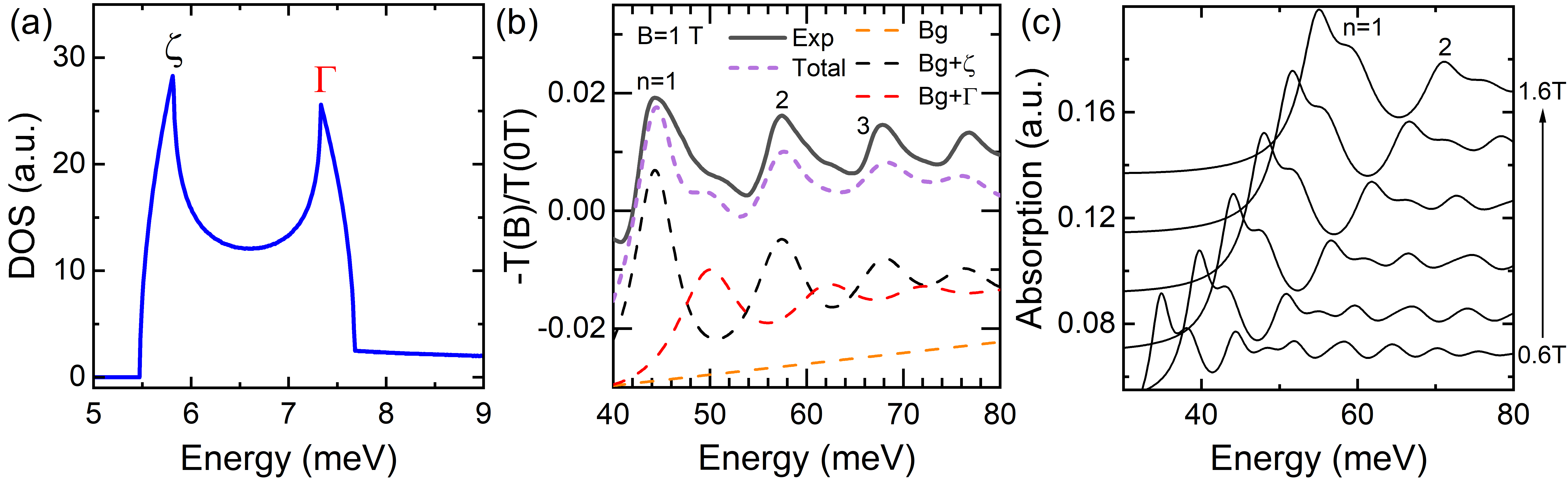}
\caption{(color online) (a) 1D DOS calculation along the $k_b$ direction based on the extracted band parameters in Table I of the main text. (b) Calculated absorption spectra (dash lines) compared to the experimental spectra (solid line) measured with $B \parallel b$-axis at 1 T. The thin (thick) dash lines show the individual (total) optical weight contributions using Eq. (S9). Specifically, the black (red) thin dash lines are the $\zeta$ ($\Gamma$) point contributions with a linear background (orange). The linewidth of each mode is set to be proportional to the energy, $\tau^{-1}_\zeta=0.04E$ and $\tau^{-1}_\Gamma=0.06E$, like that for Dirac fermions in graphene \cite{Orlita_new}. (c) Simulation of the magnetic field dependence of the two sets of LL transitions at $\zeta$ and $\Gamma$ points.}
\end{figure*}

In Fig. S3(a) (and also Fig. 3(c) of the main text), we assign the stronger mode of the doublet to the $\zeta$ point because it has lower energy. This assignment is consistent with the reconstructed low-energy electronic structure of ZrTe$_5$, similar to the violet line in Fig. 3(a) of the main text. One can further calculate the optical weight ratio of the LL transitions from the $\zeta$ and $\Gamma$ points. In Fig. S4(a), we first plot the 1D density of states (DOS) along the $k_b$ direction calculated using the extracted band parameters in Table I of the main text. The calculation is done by counting all the possible states for given energy along the $k_b$ direction within the $\pm k_BT$ energy range. The result shows a higher DOS at the $\zeta$ point than the $\Gamma$ point with a ratio of 1.15.

Next, one can calculate the absorption coefficient $\alpha$ or the real part of the optical conductivity $Re(\sigma)$, as a function of the photon energy $\hbar\omega$, using Fermi's golden rule
\begin{equation}
\alpha(\hbar \omega)\propto Re(\sigma(\hbar \omega)) \propto \frac{B}{\hbar \omega}\left| M\right|^2\frac{\tau^{-1}}{(E_f-E_i-\hbar\omega)^2+\tau^{-2}} D(\hbar\omega),\tag{S8}
\end{equation}
where $M$ is the corresponding matrix element, $D(\hbar \omega)$ is the joint DOS, $E_{f(i)}$ is the final (initial) LL energy, and $\tau^{-1}$ is the linewidth broadening. The peak height of each mode can then be estimated by $D/(\hbar \omega \tau)$, and the observed magneto-transmission spectra, $-T(B)/T(0\text{T})$, can be described as
\begin{equation}
\alpha(\hbar \omega)=\alpha_0+a\hbar\omega+A_\zeta\Sigma_\zeta \frac{1}{\hbar \omega} \frac{\tau_{\zeta}^{-1}}{(E_\zeta-\hbar\omega)^2+\tau_{\zeta}^{-2}}+A_\Gamma \Sigma_\Gamma \frac{1}{{\hbar \omega}} \frac{\tau_{\Gamma}^{-1}}{(E_\Gamma-\hbar\omega)^2+\tau_{\Gamma}^{-2}}, \tag{S9}
\end{equation}
with a linear background and two distinct sets of LL transitions from the $\zeta$ and $\Gamma$ points, respectively.  Here, the parameter $A_{\zeta(\Gamma)}$ reflects the DOS at $\zeta$ ($\Gamma$) point, and we set $A_\zeta/A_\Gamma=1.15$. The variables in this analysis are the linear background (parameter $a$) and the linewidth of each mode. Empirically, we find that our experiment can be qualitatively described by assigning linear-in-energy linewidth $\tau^{-1}_\zeta=0.04 \hbar\omega$ and $\tau^{-1}_\Gamma=0.06 \hbar\omega$, like that for Dirac fermions in graphene \cite{Orlita_new}. A side-by-side comparison between experiment and model calculation is shown in Fig. S4(b). We note that a quantitative fit to our data would require a full understanding of the scattering mechanism that determines $\tau^{-1}$. To the best of our knowledge, such information is not available for massive Dirac fermions in ZrTe$_5$ with an anisotropic electronic structure. Therefore, a quantitative fit to our data is beyond the scope of this work.

Lastly, Fig. S4(c) shows the calculated magnetic field dependence of the normalized absorption spectra based on the linewidth broadening $\tau^{-1}$ obtained in Fig. S4(b). Here, although the optical weight of each LL transition increases with magnetic field, due to the field-induced broadening, the $\Gamma$ point transitions appear merging into the neighboring $\zeta$ point transitions, becoming a shoulder-like feature in the total spectra at a higher field. Figure S4(c) qualitatively describes the experimental data in Fig. S3(a).